\documentclass[runningheads]{llncs}

\usepackage[utf8]{inputenc} 
\usepackage{hyperref}       
\usepackage{url}            
\usepackage{booktabs}       
\usepackage{amsfonts}       
\usepackage{nicefrac}       
\usepackage{microtype}      
\usepackage{amsmath,amssymb}
\usepackage{xcolor}
\usepackage{graphicx}
\usepackage[export]{adjustbox}
\usepackage{array}   
\usepackage{float}
\usepackage[ruled,vlined]{algorithm2e}
\usepackage{mathtools}
\usepackage{multirow}
\usepackage[caption=false]{subfig}

\begin{document}

\title{Enhancement of Retinal Fundus Images via Pixel Color Amplification}


\author{Alex Gaudio\inst{1}\orcidID{0000-0003-1380-6620} \and
  Asim Smailagic\inst{1}\orcidID{0000-0001-8524-997X} \and
  Aur\'{e}lio Campilho\inst{2, 3}\orcidID{0000-0002-5317-6275}}


\institute{Carnegie Mellon University, Pittsburgh PA 15213, USA \and
INESC TEC, Porto \and 
Faculty of Engineering, University of Porto
\email{agaudio@andrew.cmu.edu,asim@cs.cmu.edu,campilho@fe.up.pt}}

\renewcommand{\t}{\mathbf{t}}
\newcommand{\y}{\mathbf{y}}
\newcommand{\x}{\mathbf{x}}
\newcommand{\I}{\mathbf{I}}
\newcommand{\A}{\mathbf{A}}
\newcommand{\J}{\mathbf{J}}
\newcommand{\Z}{\mathbf{Z}}
\newcommand{\D}{\mathbf{D}}
\newcommand{\Scal}{{\mathcal{S}}}
\newcommand{\solvet}{\texttt{solve\_t}}
\newcommand{\solvetmax}{\texttt{solveMax\_t}}
\newcommand{\solvetmin}{\texttt{solveMin\_t}}
\newcommand{\solveJ}{\texttt{solve\_J}}
\newcolumntype{L}{>{$ }l<{$ }}  

\maketitle

\begin{abstract}
  We propose a pixel color amplification theory and family of enhancement methods to facilitate segmentation tasks on retinal images.  Our novel re-interpretation of the image distortion model underlying dehazing theory shows how three existing priors commonly used by the dehazing community and a novel fourth prior are related.  We utilize the theory to develop a family of enhancement methods for retinal images, including novel methods for whole image brightening and darkening.  We show a novel derivation of the Unsharp Masking algorithm.  We evaluate the enhancement methods as a pre-processing step to a challenging multi-task segmentation problem and show large increases in performance on all tasks, with Dice score increases over a no-enhancement baseline by as much as 0.491.  We provide evidence that our enhancement preprocessing is useful for unbalanced and difficult data.  We show that the enhancements can perform class balancing by composing them together.

\end{abstract}

\section{Introduction}
\begin{figure}
    \centering
    \includegraphics[width=\textwidth]{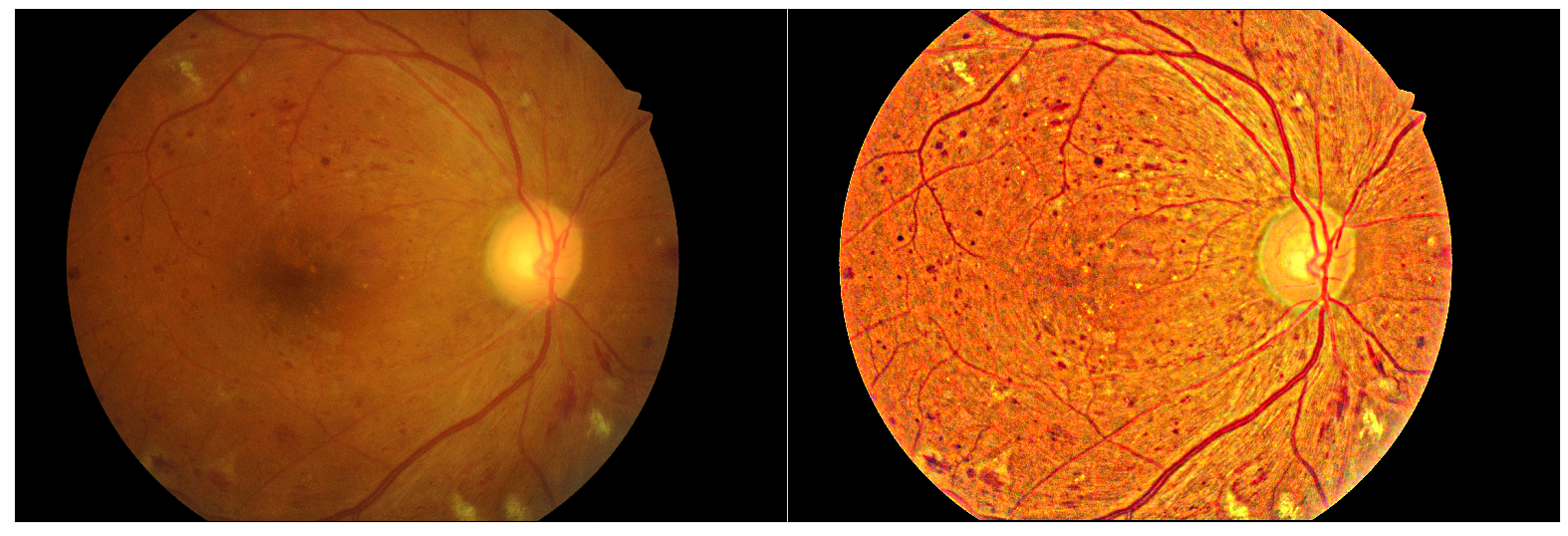}
    \caption{Comparing unmodified image (left) to our enhancement of it (right).}
    \label{fig:fig1}
\end{figure}

Image enhancement is a process of removing noise from images in order to improve performance on a future image processing task.  We consider image-to-image pre-processing methods intended to facilitate a downstream image processing task such as Diabetic Retinopathy lesion segmentation, where the goal is to identify which pixels in an image of a human retina are pathological.  In this setting, image enhancement does not in itself perform segmentation, but rather it elucidates relevant features.  Fig. \ref{fig:fig1} shows an example enhancement with our method, which transforms the color of individual pixels and enhances fine detail.

Our main contributions are to re-interpret the distortion model underlying dehazing theory as a theory of pixel color amplification.  Building on the widely known Dark Channel Prior method \cite{he_DCP}, we show a novel relationship between three previously known priors and a fourth novel prior.  We then use these four priors to develop a family of brightening and darkening methods.  Next, we show how the theory can derive the Unsharp Masking method for image sharpening.
Finally, show that the pre-processing enhancement methods improve performance of a deep network on five retinal fundus segmentation tasks.
We also open source our code for complete reproducibility \cite{ietk_github}.

\section{Related Work} \label{related_work}
Natural images are distorted by refraction of light as it travels through the transmission medium (such as air), causing modified pixel intensities in the color channels of the image.  A widely used physical theory for this distortion has traditionally been used for single image dehazing \cite{fattal_dehazing,he_DCP,srinivasa_atmosphere,tan_visibility_bad_weather}: 
\begin{align}\I(\x) = \J(\x) t(\x) + \A(1-t(\x)),\label{dehazemodel}\end{align}
where each pixel location, $\x$, in the distorted RGB image, $\I$, can be constructed as a function of the distortion-free radiance image $\J$, a grayscale transmission map image $\t$ quantifying the relative portion of the light ray coming from the observed surface in $\I(\x)$ that was not scattered (and where $t(\x) \in [0, 1] \;\forall\; \x$), and an atmosphere term, $\A$, which is typically a RGB vector that approximates the color of the uniform scattering of light.  Distortion is simply a non-negative airlight term $\A(1-t(\x))$.  We refer to \cite{fattal_dehazing} for a deeper treatment of the physics behind the theory in a dehazing context.  Obtaining a distortion free image $\J$ via this theory is typically a three step process.  Given $\I$, define an atmosphere term $\A$, solve for the transmission map $\t$, and then solve for $\J$.  We develop new insights into this theory by demonstrating ways in which it can behave as a pixel amplifier when $\t$ and $\A$ are allowed to be three channel images.

The well known Dark Channel Prior (DCP) method \cite{he_DCP,dcp_survey} addresses the dehazing task for Eq. \eqref{dehazemodel} by imposing a prior assumption on RGB images.  The assumption differentiates the noisy (hazy) image, $\I$, from its noise free (dehazed) image, $\J$.  That is, in any haze-free multi-channel region of a RGB image, at least one pixel has zero intensity in at least one channel ($\{(0,g,b),(r,0,b),(r,g,0)\}$), while a hazy region will have no pixels with zero intensity $(r>0,g>0,b>0)$.  The assumption is invalid if any channel of a distorted image is sparse or if all channels of the undistorted image are not sparse.  To quantify distortion in an image, the assumption justifies creating a fourth channel, known as the dark channel, by applying a min operator convolutionally to each region of the images $\I$ and $\J$.  
Specifically, $\tilde I^\text{dark}(\x) = \min_{c}\min_{\y\in \Omega_\I(\x)} \frac{I^{(c)}(\y)}{A^c}$, where $c$ denotes the color channel (red, green or blue) and $\Omega_\I(\x)$ is a set of pixels in $\I$ neighboring pixel $\x$.  The min operator causes $\tilde I^\text{dark}$ to lose fine detail, but an edge-preserving filter known as the guided filter \cite{he_guided_filter} restores detail $\I^\text{dark} = g(\tilde\I^\text{dark}, \I)$.  While $\J^\text{dark}(\x)$ always equals zero and therefore cancels out of the equations, $\I^\text{dark}(\x)$ is non-zero in hazy regions.  By observing that the distortion free image $\J^\text{dark}$ is entirely zero while $\I^\text{dark}$ is not entirely zero, solving Eq. \eqref{dehazemodel} for $\t$ leads to Eq. \eqref{dcp_solve_t} and then Eq. \eqref{dcp_solve_j} in Fig. \ref{fig:dcp_vs_inverted_math}.  In practice, the denominator of \eqref{dcp_solve_j} is $\max(t(\x), \epsilon)$ to avoid numerical instability or division by zero; this amounts to preserving a small amount of distortion in heavily distorted pixels. Fig. \ref{fig:dcp_vs_inverted_math} summarizes the mathematics.

\begin{figure}[tpb]
  \centering
  \small
\begin{minipage}{0.49\textwidth}\centering
  \textbf{Dark Channel Prior (Dehazing)} \\
\begin{align}
&\A = (r,g,b).\\
  &\tilde t(\x) = 1 - \min_c\min_{\y\in\Omega_{I(\x)}} \frac{I^c(\y)}{A^c}\\
  &t(\x) = \texttt{guidedFilter}(\I, \tilde t(\x)). \label{dcp_solve_t}\\
&\J(\x) = \frac{\I(\x) - \A}{\max(t(\x), \epsilon)} + \A. \label{dcp_solve_j}\\
&\implies  \J = f_\text{DCP}(\I, \A)
\end{align}
\end{minipage}%
\begin{minipage}{0.5\textwidth}\centering
  \textbf{Inverted DCP (Illumination Correction)}\\
\begin{align}
&\A = (1,1,1). \label{inv_dcpA} \\
&\J = 1-f_\text{DCP}(1-\I, \A) \label{inv_dcp}
\end{align}
\textbf{Bright Channel Prior (Exposure Correction)}\\
\begin{align}
  &\tilde t(x) = 1 - \max_c\max_{\y\in\Omega_{I(\x)}} \frac{I^c(\y)}{A^c} \\
  &t(\x) = \texttt{guidedFilter}(\I, \tilde t(\x)). \label{bcp_solve_t}\\
&\implies  \J = f_\text{BCP}(\I, \A)
\end{align}
\end{minipage}
\caption{\textbf{Left:} Dark Channel Prior (DCP) method for dehazing.  Given an (inverted) image $\I$ and atmosphere $\A$, obtain transmission map $\t$ and then recover $\J$, the undistorted image. \textbf{Top and Bottom Right:} Two priors based on inversion of the Dark Channel Prior.}
  \label{fig:dcp_vs_inverted_math}
\end{figure}

The DCP method permits various kinds of inversions.  The bright channel prior \cite{bright_channel_prior} solves for a transmission map by swapping the min operator for a max operator in Eq. \eqref{dcp_solve_t}. This prior was shown useful for exposure correction.  Fig. \ref{fig:dcp_vs_inverted_math} shows our variation of the bright channel prior based more directly on DCP mathematics and with an incorporated guided filter.  Another simple modification of the DCP method is to invert the input image $\I$ to perform illumination correction \cite{illumination_dehazing,galdran_retinex,smailagic_inverted_dehazing}.
The central idea is to invert the image, apply the dehazing equations, and then invert the dehazed result.  We demonstrate the mathematics of this inverted DCP method in Figure \ref{fig:dcp_vs_inverted_math}.  Color illumination literature requires the assumption that $\A=(1,1,1)$, meaning the image is white-balanced.  In the dehazing context, this assumption would mean the distorted pixels are too bright, but in the color illumination context, distorted pixels are too dark.  In the Methods section, we expand on the concept of brightness and darkness as pixel color amplification, show the theory supports other values of $\A$, and we also expand on the concept of inversion of Eqs. \eqref{dcp_solve_t} and \eqref{dcp_solve_j} for a wider variety of image enhancements.

\section{Methods} \label{methods}

The distortion theory Eq. \eqref{dehazemodel} is useful for image enhancement.  In section \ref{sec:amplification}, we show how the theory is a pixel color amplifier.  In section \ref{sec:invertible}, we show ways in which the theory is invertible.  We apply these properties to derive a novel prior and present a unified view of amplification under four distinct priors.  Sections \ref{sec:brightening_darkening} and \ref{sec:sharpening} apply the amplification theory to three specific enhancement methods: whole image brightening, whole image darkening and sharpening.

\subsection{The distortion theory amplifies pixel intensities} \label{sec:amplification}

We assume that $\A$, $\I$ and $\J$ share the same space of pixel intensities, so that in any given channel $c$ and pixel location $\x$, the intensities $A^c$, $I^c(\x)$ and $J^c(\x)$ can all have the same maximum or minimum value.
We can derive the simple equation $t(\x) = \frac{I^{(c)}(\x) - A^{(c)}}{J^{(c)}(\x) - A^{(c)}} \in [0, 1]$ from Eq. \eqref{dehazemodel} by noting that the distortion theory presents a linear system containing three channels.  The range of $\t$ implies the numerator and denominator must have the same sign.  For example, if $A^{(c)} \geq I^{(c)}(\x)$, then the numerator and denominator are non-positive and $J^{(c)}(\x) \leq I^{(c)}(\x) \leq A^{(c)}$.  Likewise, when $A^{(c)} \leq I^{(c)}(\x)$, the order is reversed $J^{(c)}(\x) \geq I^{(c)}(\x) \geq A^{(c)}$.  These two ordering properties show the distortion theory amplifies pixel intensities.  The key insight is that the choice of $\A$ determines how the color of each pixel in the recovered image $\J$ changes.  Models that recover $\J$ using Eq. \eqref{dehazemodel} will simply amplify color values for each pixel $\x$ in the direction $\I(\x) - \A$.

\textbf{Atmosphere controls the direction of amplification in color space.}
The atmosphere term $\A$ is traditionally a single RGB color vector with three scalar values, $A = (r,g,b)$, but it can also be an RGB image matrix.  As a RGB vector, $\A$ does not provide precise pixel level control over the amplification direction.  For instance, two pixels with the same intensity are guaranteed to change color in the same direction, even though it may be desirable for these pixels to change color in opposite directions.  Fortunately, considering $\A$ as a three channel RGB image enables precise pixel level control of the amplification direction.  It is physically valid to consider $\A$ as an image since the atmospheric light may shift color across the image, for instance due to a change in light source.  As an image, $\A$ can be chosen to define the direction of color amplification $I^c(\x)-A^c(\x)$ for each pixel and each color channel independently.

\textbf{Transmission map and Atmosphere both control the rate of amplification.}
Both the transmission map $\t$ and the magnitude of the atmosphere term $\A$ determine the amount or rate of pixel color amplification.  The effect on amplification is shown in the equation $\J = \frac{\I-\A}{\t} + \A$, where the difference $\I-\A$ controls the direction and magnitude of amplification and $\t$ affects the amount of difference to amplify.  The transmission map itself is typically a grayscale image matrix, but it can also be a scalar constant or a three channel color image.  Each value $t(\x) \in [0,1]$ is a mixing coefficient specifying what proportion of the signal is not distorted.  When $t(\x)=1$, there is no distortion; the distorted pixel $\I(\x)$ and corresponding undistorted pixel $\J(\x)$ are the same since $\I(\x) = \J(\x) + 0$.  As $t(\x)$ approaches zero, the distortion caused by the difference between the distorted image $\I$ and the atmosphere increases.

\subsection{Amplification under inversion}
\label{sec:invertible}

The distortion theory supports several kinds of inversion.  The equations \eqref{dcp_solve_t} and \eqref{dcp_solve_j} are invertible.  The input image $\I$ can also undergo invertible transformations.  We prove these inversion properties and show why they are useful.

\textbf{Inverting Eq. \eqref{dcp_solve_t} results in a novel DCP-based prior.} We discussed in Related Work three distinct priors that provide a transmission map: the traditional DCP approach with Eq. \eqref{dcp_solve_t}; the bright channel prior in Eq. \eqref{bcp_solve_t}; and color illumination via Eq. \eqref{inv_dcp}.  Bright channel prior and color illumination respectively perform two types of inversion; the former changes the min operator to a max operator while the latter inverts the image $1-\I$.  Combining these two inversion techniques results in a novel fourth prior.  In Table \ref{table:inversion}, we show the four transmission maps.
We show that each prior has a solution using either the min or max operator, which is apparent by the following two identities:

\begin{align}
  \solvet(\I,\A) = 1 - \min_c\min_{y\in\Omega_{I(\x)}} \frac{I^c(\y)}{A^c} &\equiv \max_c\max_{y\in\Omega_{I(\x)}} \frac{1-I^c(\y)}{A^c}\\
  \solvet(\I,\A) = 1 - \max_c\max_{y\in\Omega_{I(\x)}} \frac{I^c(\y)}{A^c} &\equiv \min_c\min_{y\in\Omega_{I(\x)}} \frac{1-I^c(\y)}{A^c}
\end{align}

The unified view of these four priors in Table \ref{table:inversion} provides a novel insight into how they are related.  In particular, the table provides proof that the Color Illumination Prior and Bright Channel Prior are inversely equivalent and utilize statistics of the maximum pixel values across channels. Similarly, DCP and our prior are also inversely equivalent and utilize statistics of the minimum pixel values across channels.  This unified view distinguishes between weak and strong amplification, and amplification of bright and dark pixel neighborhoods.

In Figure \ref{fig:amplification_t}, we visualize these four transmission maps to demonstrate how they collectively perform strong or weak amplification of bright or dark regions of the input image.
In this paper, we set $\A=\mathbf{1}$ when solving for $\t$.  Any choice of $A^c \in (0, 1]$ is valid, and when all $A^c$ are equal, smaller values of $\A$ are guaranteed to amplify the differences between these properties further.

\begin{table}[H]
  \small
  \centering
  \caption{Four transmission maps derived from variations of Eq. \eqref{dcp_solve_t}.
  For clear notation, we used the vectorized functions {\scriptsize$\t = \solvetmin(\I, \A) = 1 - \min_c\min_{\y\in\Omega_{I(\x)}} \frac{I^c(\y)}{A^c}$} and {\scriptsize $\t = \solvetmax(\I, \A) = 1 - \max_c\max_{\y\in\Omega_{I(\x)}} \frac{I^c(\y)}{A^c}$}. }
  \label{table:inversion}
  \begin{tabular}{l|L|L}
    \hline
    &\textbf{\scriptsize Amplify Dark Areas}&\textbf{\scriptsize Amplify Bright Areas}\\
    \hline
    \shortstack{\textbf{\scriptsize Weak Amplification}}
      & \shortstack{\\ $\begin{aligned} &\solvetmin(1-\I, \A=1) \\[-5pt] 1-&\solvetmax(\I, \A=1)   \end{aligned}$ \\ \text{Color Illumination Prior}}
      & \shortstack{\\ $\begin{aligned} &\solvetmin(\I, \A=1)   \\[-5pt] 1-&\solvetmax(1-\I, \A=1) \end{aligned}$ \\ \text{Standard Dark Channel Prior}}\\
    \hline
    \shortstack{\textbf{\scriptsize Strong Amplification}}
      & \shortstack{\\ $\begin{aligned} 1-&\solvetmin(\I, \A=1)   \\[-5pt] &\solvetmax(1-\I, \A=1) \end{aligned}$ \\ \text{Our novel prior}}
      & \shortstack{\\ $\begin{aligned} 1-&\solvetmin(1-\I, \A=1) \\[-5pt] &\solvetmax(\I, \A=1)   \end{aligned}$ \\ \text{Bright Channel Prior}}\\
    \hline
  \end{tabular}
\begin{figure}[H]
  \centering
  \includegraphics[width=0.40\linewidth,valign=t]{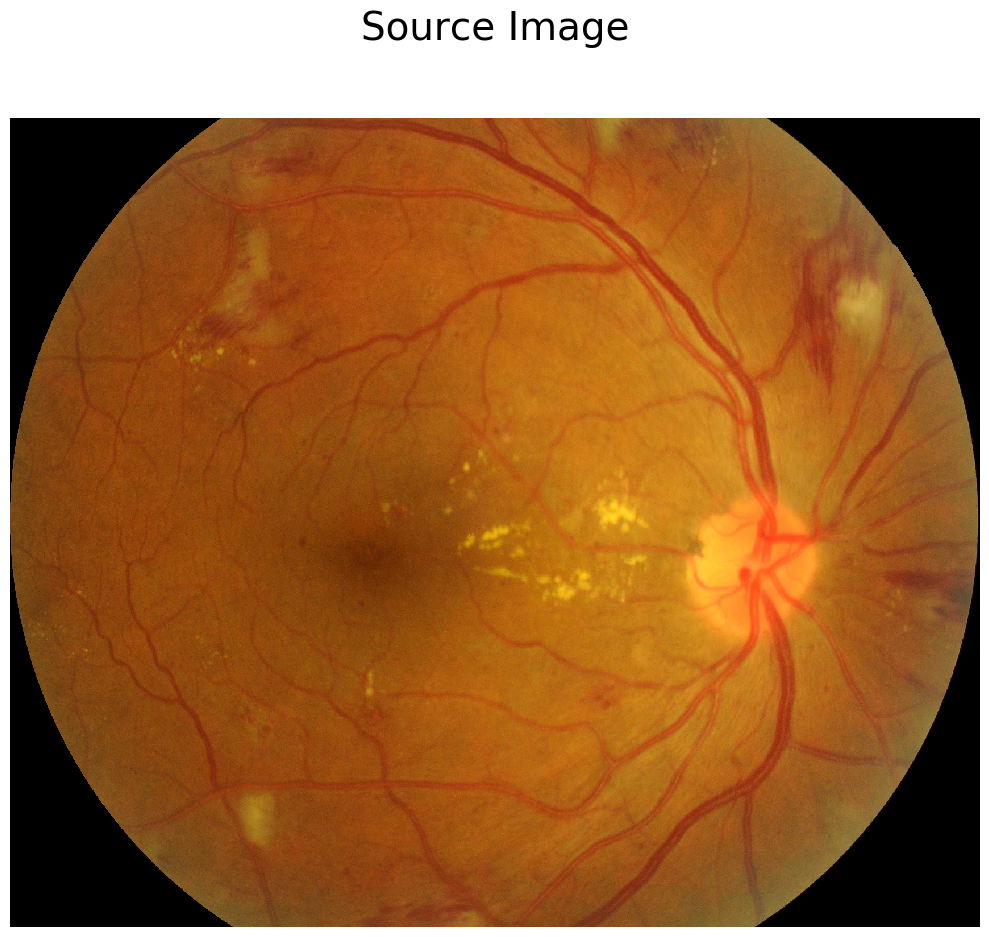}
  \includegraphics[width=0.40\linewidth,valign=t]{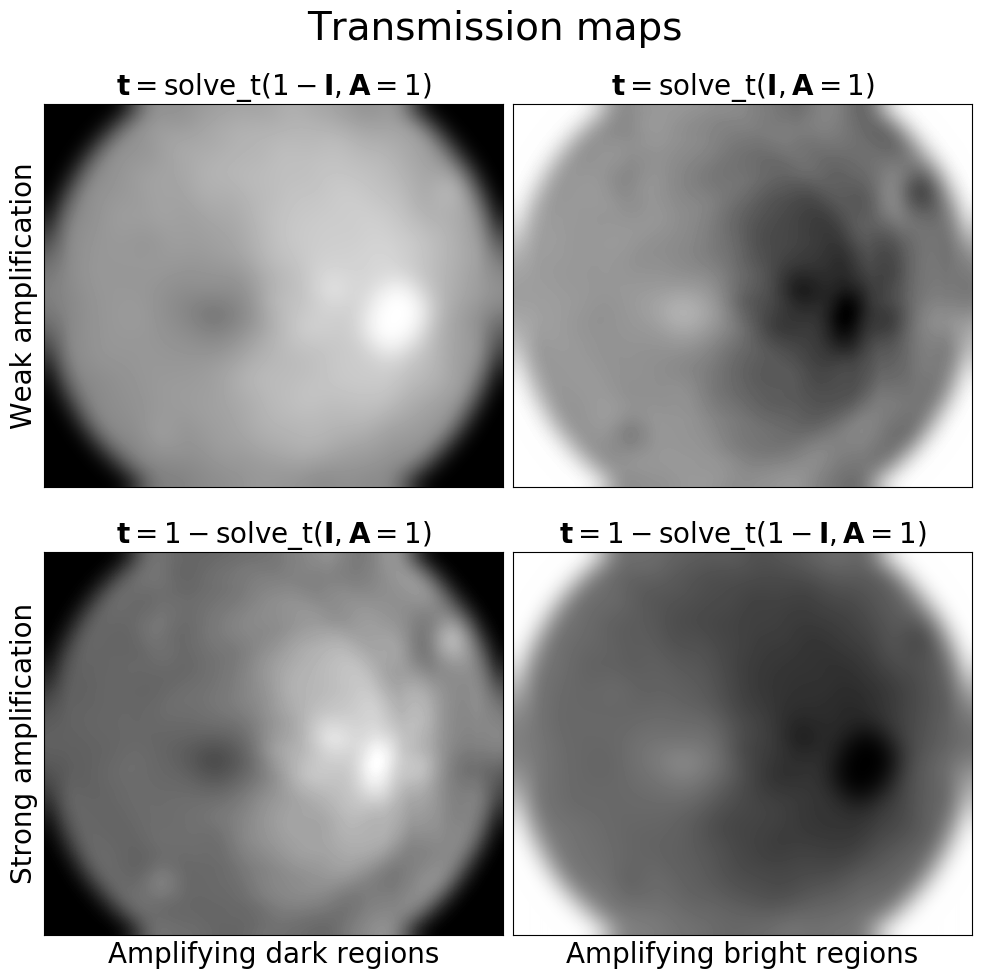}
  \caption{The transmission maps (right) obtained from source image (left) selectively amplify bright or dark regions.  Dark pixels correspond to a larger amount of amplification.
    We randomly sample a retinal fundus image from the IDRiD dataset (see Section \ref{IDRiD}).  We set the blue channel to all ones when computing the transmission map for the top right and bottom left maps because the min values of the blue channel in retinal fundus images are noisy.}
  \label{fig:amplification_t}
\end{figure}
\end{table}

\textbf{Inverting Eq. \eqref{dcp_solve_j} motivates brightening and darkening.} Given an image $\I$, transmission map $\t$ and an atmosphere $\A$, solving for the recovered image $\J$ with Eq. \eqref{dcp_solve_j} can be computed two equivalent ways, as we demonstrate by the following identity:

\begin{align}
  \J = \texttt{solve\_J}(\I, \t, \A) \equiv 1-\texttt{solve\_J}(1-\I, \t, 1-\A) \label{eq:identity_dcp_solve_j}
\end{align}

The proof is by simplification of $\frac{\I-\A}{\t} + \A = 1-\left(\frac{ (1-\I)-(1-\A) }{\t}+(1-\A)\right)$.  It implies the space of possible atmospheric light values, which is bounded in [0,1], is symmetric under inversion.

We next prove that solving for $\J$ via the color illumination method \cite{illumination_dehazing,galdran_retinex,smailagic_inverted_dehazing} is equivalent to direct attenuation $\J=\frac{\I}{\t}$, a fact that was not clear in prior work.  As we noted in Eq. \eqref{inv_dcp}, color illumination solves $\J = 1-\frac{(1-\I)-\A}{\t}+\A$ under the required assumption that $\A=\mathbf{1}$.  We can also write the atmosphere as $\A=1-\mathbf{0}$.  Then, the right hand side of \eqref{eq:identity_dcp_solve_j} leads to $\J=1-\solveJ(1-\I, \t, \A=1-\mathbf{0}) = \frac{\I-\mathbf{0}}{\t} + \mathbf{0}$.  Therefore, color illumination actually performs whole image brightening with the atmosphere $\A=(0,0,0)$ even though the transmission map uses a white-balanced image assumption that $\A=(1,1,1)$. Both this proof and the invertibility property Eq. \eqref{eq:identity_dcp_solve_j} motivate Section \ref{sec:brightening_darkening} where we perform brightening and darkening with all priors in Table \ref{table:inversion}.

\subsubsection{Application to Whole Image Brightening and Darkening.}\label{sec:brightening_darkening}

\begin{figure}[!htbp]
  \centering
  \includegraphics[width=0.49\linewidth]{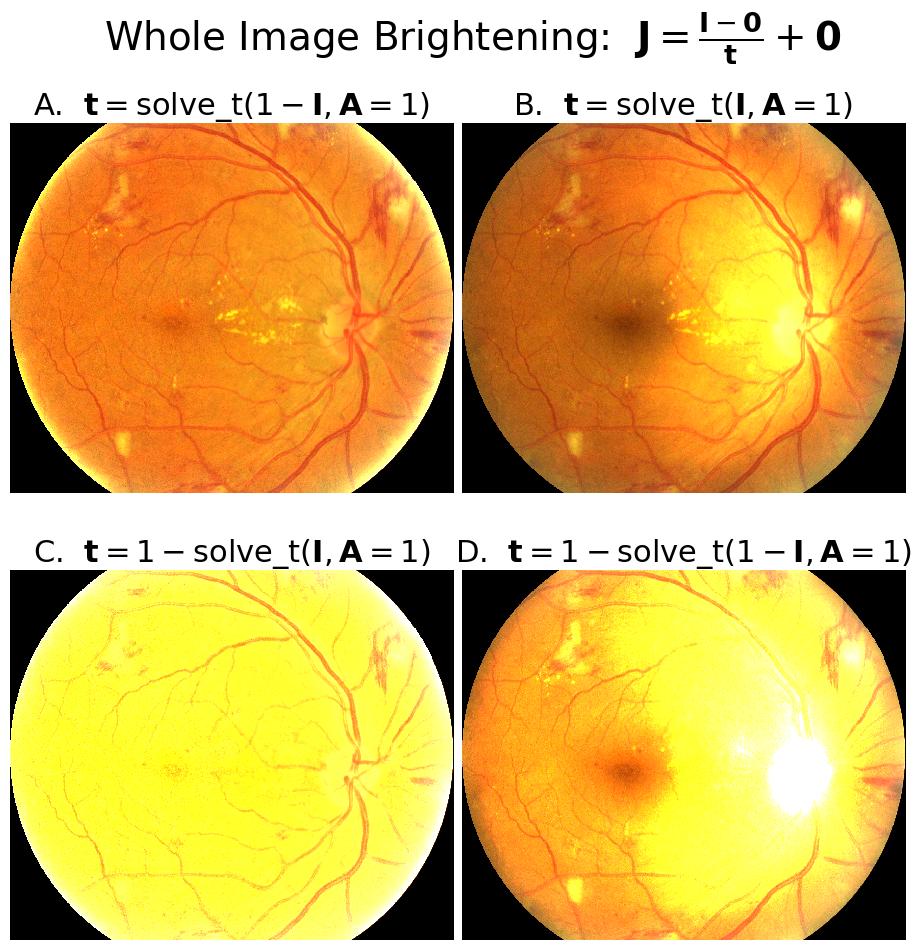}
  \includegraphics[width=0.49\linewidth]{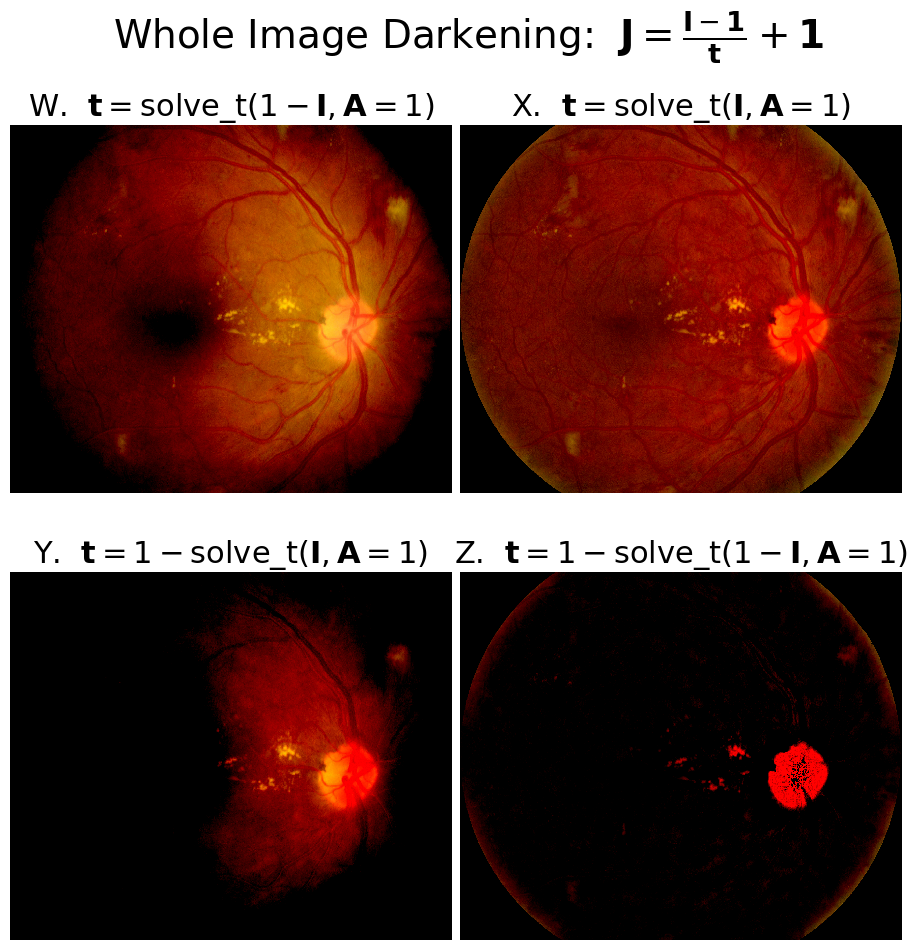}
  \caption{Whole image Brightening (Left) and Darkening (Right) using the corresponding four transmission maps in Fig. \ref{fig:amplification_t}.  Note that $\A=\mathbf{1}$ when solving for $\t$, but $\A=\mathbf{0}$ or $\A=\mathbf{1}$, respectively, for brightening or darkening.}
  \label{fig:amplification_bd}
\end{figure}

Brightening versus darkening of colors is a matter of choosing an amplification direction.  Extremal choices of the atmosphere term $\A$ result in brightening or darkening of all pixels in the image.  For instance, $\A=(1,1,1)$ guarantees for each pixel $\x$ that the recovered color $\J(\x)$ is darker than the distorted color $\I(\x)$ since $\J \leq \I \leq \A$, while $\A=(0,0,0)$ guarantees image brightening $\J \geq \I \geq \A$.   More generally, any $\A$ satisfying $1 \geq A^c \geq \max_\x I^c(\x)$ performs whole image brightening and any $\A$ satisfying $0 \leq A^c \leq \min_\x I^c(\x)$ performs whole image darkening.
We utilize the four distinct transmission maps from Table \ref{table:inversion} to perform brightening $\A=\mathbf{0}$ or darkening $\A=\mathbf{1}$, resulting in eight kinds of amplification.  We visualize these maps and corresponding brightening and darkening techniques applied to retinal fundus images in Fig. \ref{fig:amplification_bd}.  Our application of the Bright Channel Prior and Color Illumination Prior for whole image darkening is novel. Utilizing our prior for brightening and darkening is also novel.

\begin{figure}[htbp]
  \centering
  \begin{figure}[H]
    \centering
  \includegraphics[width=.95\linewidth]{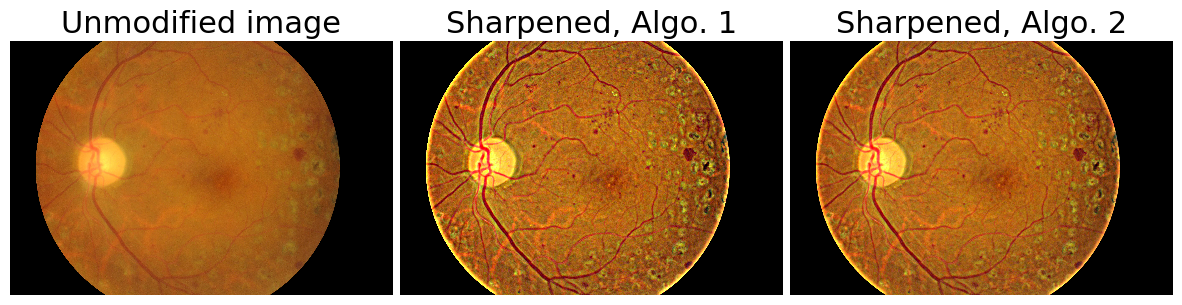}
  \caption{Sharpening a retinal fundus image with Algo. \ref{algo:sharpen} (middle) and \ref{algo:sharpen_laplace} (right).  Image randomly sampled from IDRiD training dataset (described in Sec. \ref{IDRiD}).}
  \label{fig:sharpening_fundus}
  \end{figure}
\SetKwInput{KwInput}{Input} 
\centering
\begin{minipage}[t]{0.45\textwidth}
  \small
  \begin{algorithm}[H]
\small
  \SetAlgoLined
  \KwInput{$\I$ \quad\small(input fundus image)}
  \KwResult{$\J$  \quad\small(sharpened image)}
  $\A = \texttt{blur}(\I, \texttt{blur\_radius})$\;
  $\t = 0.15$\;

  \eIf{$\min(\text{img\_width}, \text{img\_height}) > 1500$}{
    $\J = \texttt{guidedFilter}(\texttt{guide}=\I, \texttt{src}=\frac{\I-\A}{\t} + \A)$\;
  }{
    $\J = \frac{\I-\A}{\t} + \A$\;
  }
  \caption{Image Sharpening, simple}
  \label{algo:sharpen}
\end{algorithm}
\end{minipage}
\begin{minipage}[t]{0.49\textwidth}
\begin{algorithm}[H]
  \small
  \SetAlgoLined
  \KwInput{$\I$ \quad\small(input fundus image)}
  \KwResult{$\J$  \quad\small(sharpened image)}
  $\widetilde \t = \textbf{Algo\_1}($
  $\qquad\texttt{morphological\_laplace}(\I, (2,2,1)))$\;
$\widetilde \t = 1-\frac{\widetilde \t - \min(\widetilde \t)}{\max(\widetilde \t) - \min( \widetilde \t )}$\;
  $\epsilon = \max(10^{-8}, \frac{\min(\widetilde\t)}{2})$\;
  $\t = \texttt{elementwise\_max}(\widetilde \t, \epsilon)$\;
  $\J = \textbf{Algo\_1}(\I,\t=\t)$\;
  \caption{Image Sharpening, complex}
  \label{algo:sharpen_laplace}
\end{algorithm}
\end{minipage}
\begin{figure}[H]
  \centering
  \includegraphics[width=0.49\linewidth]{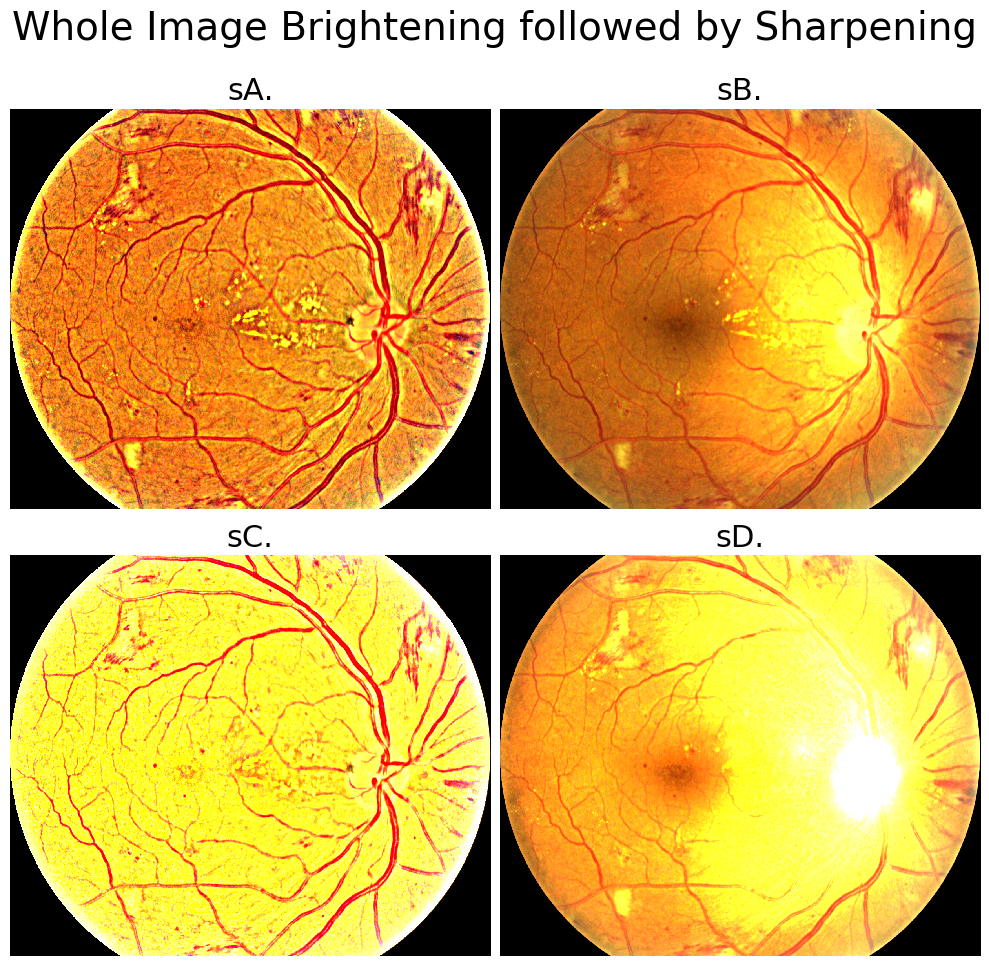}
  \includegraphics[width=0.49\linewidth]{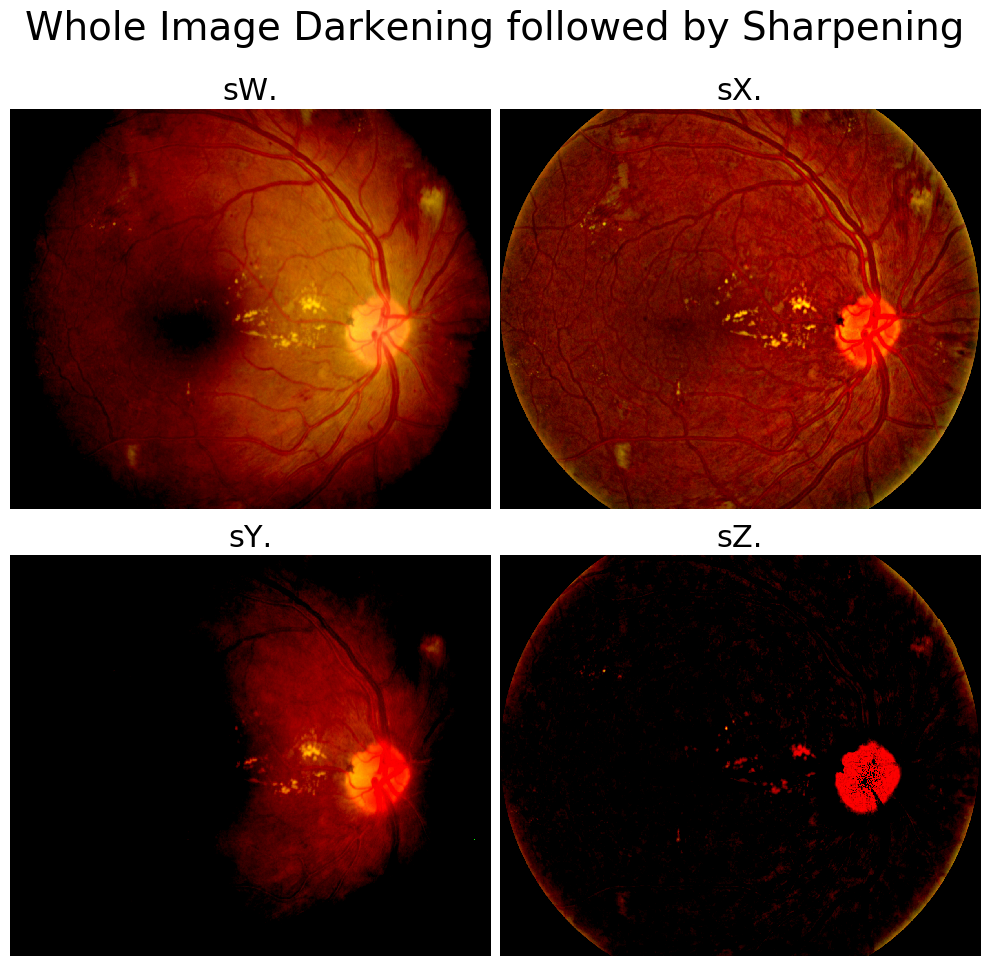}
  \caption{The result of sharpening each image in Fig. \ref{fig:amplification_bd} using Algo. \ref{algo:sharpen_laplace}.}
  \label{fig:amplification_bd_sharpen}
\end{figure}
\end{figure}
\subsubsection{Application to Image Sharpening.}
\label{sec:sharpening}
We show a novel connection between dehazing theory and \textit{unsharp masking}, a deblurring method and standard image sharpening technique that amplifies fine detail \cite{book_with_unsharp_mask_algo_p_357}.  Consider $\A$ as a three channel image obtained by applying a non-linear blur operator to $\I$, $\A = \text{blurry}(\I)$.  Solving Eq. $\eqref{dehazemodel}$ for $\J$ gives $\J = \frac{1}{\t}\I - \frac{(1-\t)}{\t}\A$.  Since each scalar value $t(\x)$ is in $[0,1]$, we can represent the fraction $t(\x) = \frac{1}{u(x)}$.  Substituting, we have the simplified matrix form $\J = \mathbf{u}\circ \I - (\mathbf{u}-1)\circ\text{blurry}(\I)$ where the $\circ$ operator denotes element-wise multiplication with broadcasting across channels. 
This form is precisely \textit{unsharp masking}, where $\mathbf{u}$ is either a constant,
or $\mathbf{u}$ is a 1-channel image matrix determining how much to sharpen each pixel.  The matrix form of $\mathbf{u}$ is known as locally adaptive unsharp masking.  Thus, we show the distortion theory in Eq. \eqref{dehazemodel} is equivalent to image sharpening by choosing $\A$ to be a blurred version of the original input image.

We present two sharpening algorithms, Algo. \ref{algo:sharpen} and \ref{algo:sharpen_laplace}, and show their respective outputs in Fig. \ref{fig:sharpening_fundus}.  Sharpening amplifies differences between an image and a blurry version of itself.  In unevenly illuminated images, the dark or bright regions may saturate to zero or one respectively.  Therefore, the use of a scalar transmission map (Algo. \ref{algo:sharpen}), where all pixels are amplified, implies that the input image should ideally have even illumination.  The optional guided filter in the last step provides edge preserving smoothing and helps to minimize speckle noise, but can cause too much blurring on small images, hence the if condition.

Algo. \ref{algo:sharpen_laplace} selectively amplifies only the regions that have an edge.  Edges are found by deriving a three channel transmission map from a Laplacian filter applied to a morphologically smoothed fundus image.  We enhance edges by recursively sharpening the Laplace transmission map under the theory.  Fig. \ref{fig:amplification_bd_sharpen} shows the results of sharpening each image in Fig. \ref{fig:amplification_bd} with Algo. \ref{algo:sharpen_laplace}.

\section{Experiments} \label{experiments}

Our primary hypothesis is that enhancement facilitates a model's ability to learn retinal image segmentation tasks.  We introduce a multi-task dataset and describe our deep network implementation.

\subsection{Datasets} 

\label{IDRiD}
The \textbf{Indian Diabetic Retinopathy Dataset (IDRiD)} \cite{IDRiD} contains 81 retinal fundus images for segmentation, with a train-test split of 54:27 images.  Each image is 4288x2848 pixels. Each pixel has five binary labels for presence of: Microaneurysms (MA), Hemorrhages (HE), Hard Exudates (EX), Soft Exudates (SE) and Optic Disc (OD).  Only 53:27 and 26:14 images present HE and SE, respectively. Table \ref{table:IDRiD} shows the fraction of positive pixels per category is unbalanced both across categories (left columns) and within categories (right columns).

\begin{table}
  \centering
  \small
  \begin{minipage}[t]{.49\linewidth}
    \caption{IDRiD Dataset,\newline An Unbalanced Class Distribution.}
\begin{tabular}{l|rr|rr}
\toprule
        & \multicolumn{2}{l}{\textbf{Pos/$\sum$Pos}} & \multicolumn{2}{l}{\textbf{Pos/(Pos+Neg)}} \\
\scriptsize Category&        \scriptsize Train &   \scriptsize Test &         \scriptsize Train &   \scriptsize Test \\
\midrule
     MA &        0.027 &  0.024 &         0.0007 &  0.0003\\
     HE &        0.253 &  0.256 &         0.0066 &  0.0036\\
     EX &        0.207 &  0.261 &         0.0054 &  0.0036\\
     SE &        0.049 &  0.043 &         0.0013 &  0.0006\\
     OD &        0.464 &  0.416 &         0.0120 &  0.0058\\
\bottomrule
\end{tabular}

    \label{table:IDRiD}
    \vspace{.5cm}
    \caption{{Competing Method Results,}\newline{Best Per Category of A, B, D, or X.}}
    \begin{tabular}{lll}
    \toprule
       &          &   Dice (delta) \\
    \textbf{Task} & \textbf{Method} &                \\
    \midrule
    \multirow{1}{*}{\textbf{EX}} & \textbf{A} &  0.496 (0.175) \\
    \cline{1-3}
    \multirow{1}{*}{\textbf{HE}} & \textbf{A} &  0.102 (0.102) \\
    \cline{1-3}
    \multirow{1}{*}{\textbf{MA}} & \textbf{A} &  0.122 (0.122) \\
    \cline{1-3}
    \multirow{1}{*}{\textbf{OD}} & \textbf{A} &  0.849 (0.332) \\
    \cline{1-3}
    \multirow{1}{*}{\textbf{SE}} & \textbf{A} &  0.423 (0.264) \\
    \bottomrule
    \end{tabular}
    \label{table:ADX}
  \end{minipage}%
  \begin{minipage}[t]{.49\linewidth}
    \caption{{Main Results,\newline Pre-processing Yields Large Improvements.}}
  \begin{tabular}{|lll}
\toprule
   &            &   Dice (delta) \\
\textbf{Task} & \textbf{Method} &                \\
\midrule
\multirow{2}{*}{\textbf{EX}} & \textbf{avg4:sA+sC+sX+sZ} &  0.728 (0.407) \\
   & \textbf{avg2:sA+sZ} &  0.615 (0.295) \\
\cline{1-3}
\multirow{2}{*}{\textbf{HE}} & \textbf{avg3:sA+sC+sX} &  0.491 (0.491) \\
   & \textbf{avg3:sB+sC+sX} &  0.368 (0.368) \\
\cline{1-3}
\multirow{2}{*}{\textbf{MA}} & \textbf{avg4:A+B+C+X} &  0.251 (0.251) \\
   & \textbf{avg2:sB+sX} &  0.219 (0.219) \\
\cline{1-3}
\multirow{2}{*}{\textbf{OD}} & \textbf{avg4:sA+sC+sX+sZ} &  0.876 (0.359) \\
   & \textbf{avg2:sA+sZ} &  0.860 (0.343) \\
\cline{1-3}
\multirow{2}{*}{\textbf{SE}} & \textbf{avg4:sA+sC+sX+sZ} &  0.491 (0.332) \\
   & \textbf{avg3:B+C+X} &  0.481 (0.322) \\
\bottomrule
\end{tabular}

  \label{table:idrid_top}
\end{minipage}
\end{table}

\subsubsection{Blackbox Evaluation: Does an enhancement method improve performance?}
\label{sec:eval_blackbox}

\newcommand{\w}{\mathbf{w}}
We implement and train a standard U-Net model \cite{unet_ronneberger} and evaluate change in performance via the Dice coefficient.  We apply this model simultaneously to five segmentation tasks (MA, HE, SE, EX, OD) on the IDRiD dataset; the model has five corresponding output channels.  We use a binary cross entropy loss summed over all pixels and output channels.  We apply task balancing weights to ensure equal contribution of positive pixels to the loss.  The weights are computed via $\frac{\max_i w_i}{\w}$, where the vector $\w$ contains counts of positive pixels across all training images for each of the 5 task categories (see left column of Table \ref{table:IDRiD}).  Without the weighting, the model did not learn to segment MA, HE, and EX even with our enhancements.  For the purpose of our experiment, we show in the results that this weighting is suboptimal as it does not balance bright and dark categories.  The Adam Optimizer has a learning rate 0.001 and weight decay 0.0001.  We also applied the following pre-processing: center crop the fundus to minimize background, resize to (512x512) pixels, apply the pre-processing enhancement method (independent variable), clip pixel values into [0,1], randomly rotate and flip.  Rotations and flipping were only applied on the training set; we excluded them from validation and test sets.
We randomly hold out two training images as the validation set in order to apply early stopping with a patience of 30 epochs.  We evaluate test set segmentation performance with the S{\o}rensen-Dice coefficient, which is commonly used for medical image segmentation.  

\subsection{Pre-Processing Enhancement Methods for Retinal Fundus Images}

We combine the brightening, darkening and sharpening methods together and perform an ablation study.  We assign the eight methods in Fig. \ref{fig:amplification_bd} a letter.  The brightening methods, from top left to bottom right are A,B,C,D.  The corresponding darkening methods are W,X,Y,Z.  We also apply sharpening via Algo. \ref{algo:sharpen_laplace}.  Combined methods assume the following notation: $A+X$ is the average of $A$ and $X$, which is then sharpened; $sA+sX$ is the average of sharpened $A$ with sharpened $X$.  A standalone letter $X$ is a sharpened $X$.  All methods have the same hyperparameters, which we chose using a subset of IDRiD training images.  When solving for $t$, the size of the neighborhood $\Omega$ is (5x5); the guided filter for $t$ has radius=100, $\epsilon=1e^{-8}$.  When solving for $J$, the $\max$ operator in the denominator is $\max\langle \min(t)/2, 1e-8\rangle$.  For sharpening, we blur using a guided filter (radius=30, $\epsilon = 1e^{-8}$), and we do not use a guided filter to denoise as the images are previously resized to (512x512).

\section{Results} \label{results}

\subsection{Our pre-processing enhancement methods significantly improve performance on all tasks.}

We show the top two models with highest test performance in each category in Table \ref{table:idrid_top}.  The delta values show the pre-processing enhancement methods significantly improve performance over the no-enhancement (identity function) baseline for all tasks, underscoring the value of our theory and methods.

\textbf{Enhancement improves detection of rare classes.} The smallest delta improvement in Table \ref{table:idrid_top} is 0.219 for MA, the rarest category across and within categories (as shown in Table \ref{table:IDRiD}).  Our smallest improvement is a large increase considering the largest possible Dice score is one.

\textbf{Enhancement can be class balancing.} The IDRiD results support the primary hypothesis that enhancement makes the segmentation task easier.  The delta values show the baseline identity model did not learn to segment MA or HE.  Indeed, during the implementation, we initially found that the model learned to segment only the optic disc (OD).  Of the categories, OD has the most extremal intensities (brightest) and is typically the largest feature by pixel count in a fundus image. In our multi-task setting, the gradients from other tasks were therefore overshadowed by OD gradients.  After we implemented a category balancing weight, the no-enhancement baseline model was still unable to learn MA and HE.  As an explanation for this phenomenon, we observe that EX, SE and OD are bright features while MA and HE are dark features.  Considering the class balancing weights, the bright features outnumber the dark features three to two.  This need to carefully weigh the loss function suggests differences in color intensity values \textit{cause} differences in performance.  It is therefore particularly interesting that the enhancement methods were able to learn despite also being subject to these issues.  In fact, we can observe that the best enhancements in the table incorporate the Z method, which performs a strong darkening of bright regions.  We interpret this result as strong evidence that our enhancement methods make the segmentation task easier, and in fact, that they can be used as a form of class balancing by image color augmentation.

\begin{figure}[htpb]
    \centering
    \includegraphics[width=.95\textwidth]{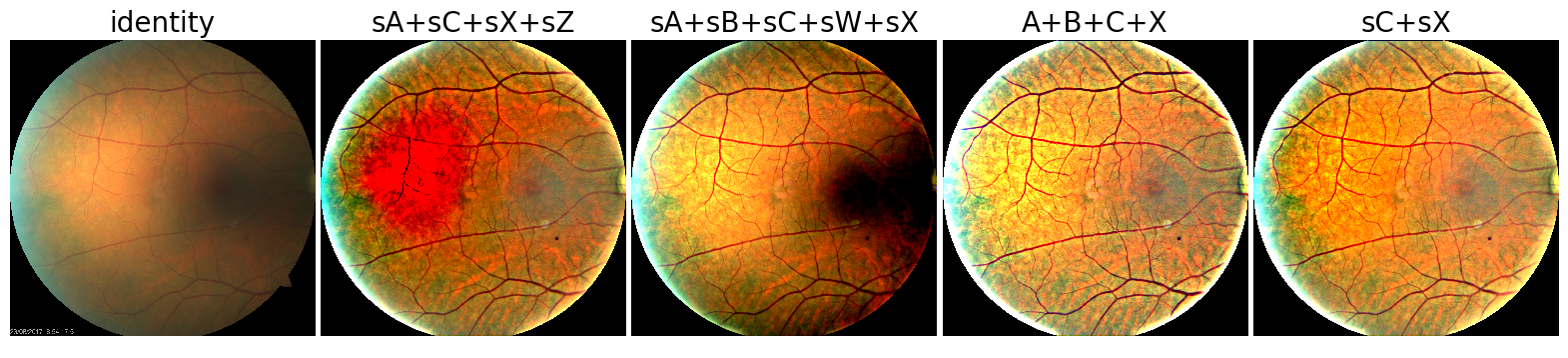}
    \includegraphics[width=.95\textwidth]{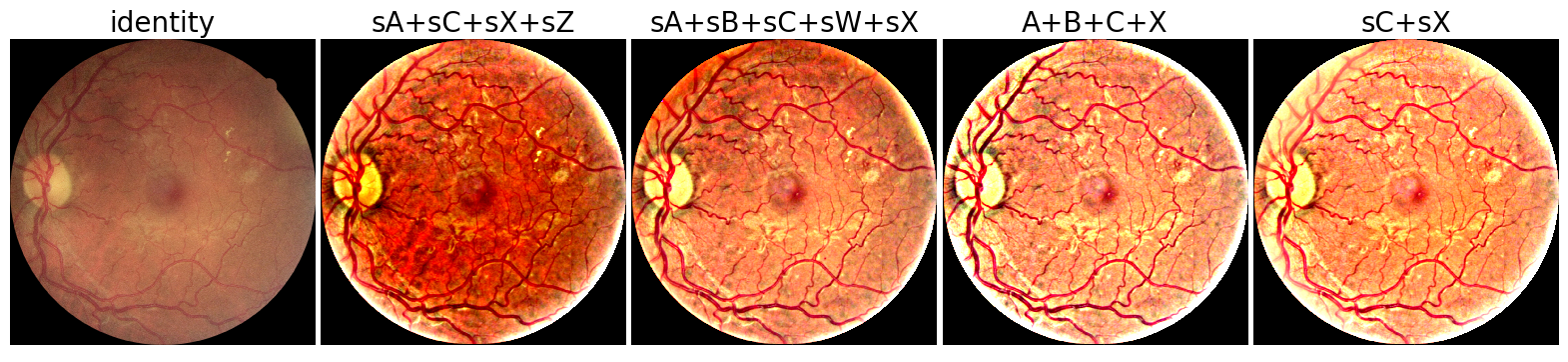}
    \includegraphics[width=.95\textwidth]{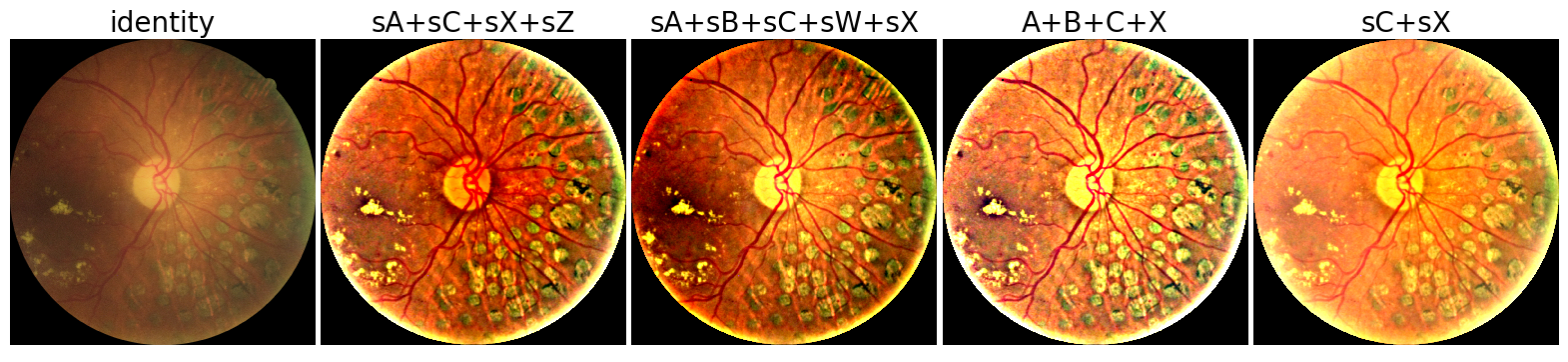}
    \includegraphics[width=.95\textwidth]{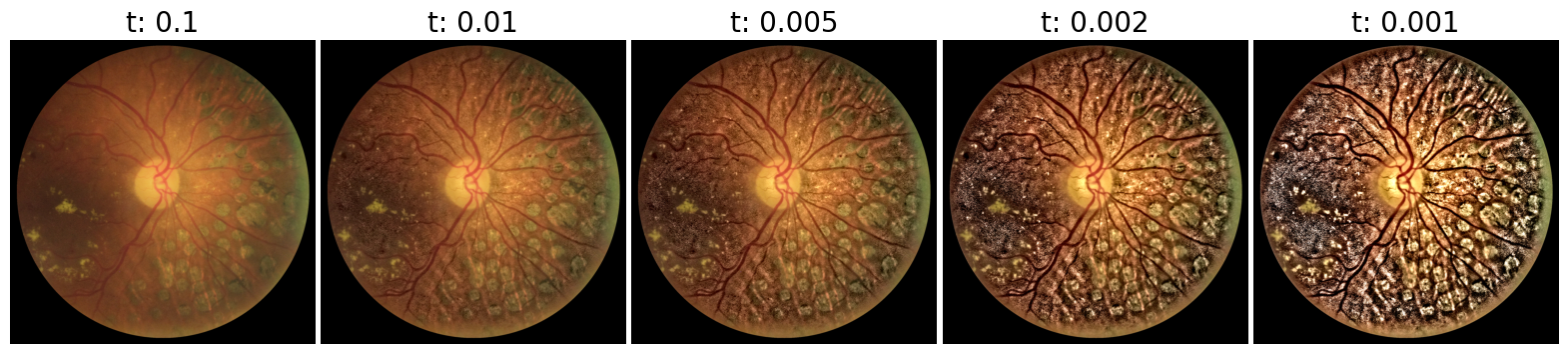}
    \includegraphics[width=.95\textwidth]{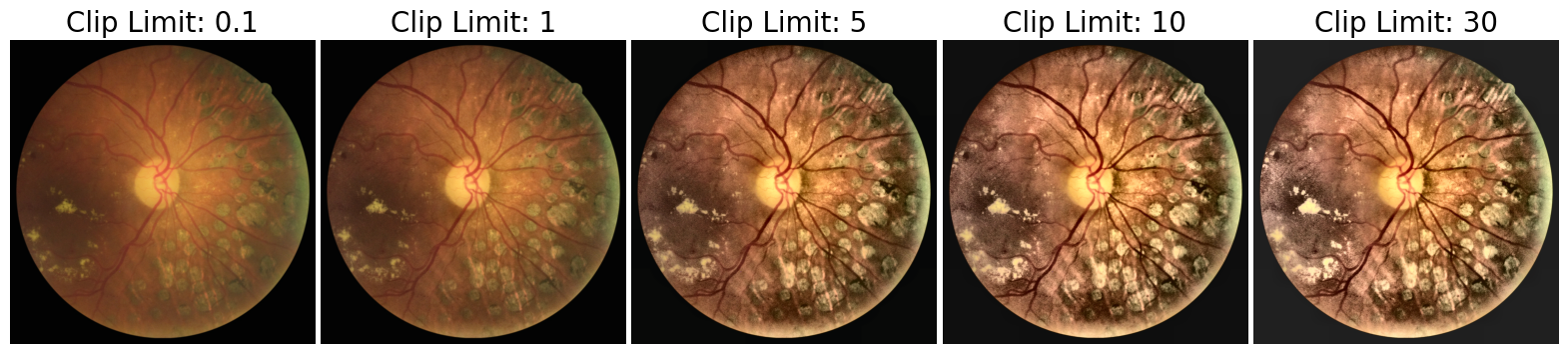}
    \caption{Visualization of our enhancement methods.  Each row is an image. Each column is an enhancement method. Last two rows compare our Algo. \ref{algo:sharpen} with CLAHE.}
    \label{fig:qualitative}
\end{figure}

\subsection{Comparison to existing work}
The methods $A, D, X$ and arguably $B$ correspond to existing work and were visualized (with sharpening) in Fig. \ref{fig:amplification_bd_sharpen}.  The $A$ method outperforms $B$, $D$ and $X$ on all tasks.  Its values, reported in Table \ref{table:ADX}, are substantially lower than the values in Table \ref{table:idrid_top}. We attribute the low scores to our intentional category imbalance.

Contrast Limited Adaptive Histogram Equalization (CLAHE) applied to the luminance channel of LAB colorspace is useful for retinal fundus enhancement \cite{clahe_competing_method}.  We compare it to Algo. \ref{algo:sharpen} in bottom rows of Fig. \ref{fig:qualitative}, using the LAB conversion for both methods.  We observe that CLAHE preserves less detail, and both methods overemphasize uneven illumination.  CLAHE is faster to compute and could serve as a simple drop-in replacement, with clip limit as a proxy for the scalar $t$.

\subsection{Qualitative Analysis}
\label{sec:results_qualitative}

We visualize a subset of our image enhancement methods in the top three rows of Fig. \ref{fig:qualitative}.  Each row presents a different fundus image from a private collection.  We observe that the input images are difficult to see and appear to have little detail, while the enhanced images are quite colorful and very detailed.  The halo effect around the fundus is caused by the guided filter (with $\epsilon=1e^{-8}$) rather than the theory.  The differences in bright and dark regions across each row provide intuitive sense of how averaging the models (Fig. \ref{fig:amplification_bd} and \ref{fig:amplification_bd_sharpen}) can yield a variety of different colorings.

\section{Conclusion}

In this paper, we re-interpret a theory of image distortion as pixel color amplification and utilize the theory to develop a family of enhancement methods for retinal fundus images.  We expose a relationship between three existing priors commonly used for image dehazing with a fourth novel prior.  We apply our theory to whole image brightening and darkening, resulting in eight enhancement methods, five of which are also novel (methods B, C, W, Y, and Z).  We also show a derivation of the Unsharp Masking algorithm for image sharpening and develop a sharpening algorithm for retinal fundus images.  Finally, we evaluate our enhancement methods as pre-processing steps for multi-task deep network retinal fundus image segmentation.  We show the enhancement methods give strong improvements and can perform class balancing.  Our pixel color amplification theory applied to retinal fundus images yields a variety of rich and colorful enhancements, as shown by our compositions of methods A-D and W-Z, and the theory shows great promise for wider adoption by the community.

\section*{Acknowledgements}

We thank Dr. Alexander R. Gaudio, a retinal specialist and expert in degenerative retinal diseases, for his positive feedback and education of fundus images.

Supported in part by the National Funds through the
Funda\c{c}\~{a}o para a Ci\^{e}ncia e a Tecnologia within under Project CMUPERI/TIC/0028/2014.

\small
\bibliographystyle{splncs04}
\bibliography{ref}

\begin{thebibliography}{10}
\providecommand{\url}[1]{\texttt{#1}}
\providecommand{\urlprefix}{URL }
\providecommand{\doi}[1]{https://doi.org/#1}

\bibitem{clahe_competing_method}
Cao, L., Li, H., Zhang, Y.: Retinal image enhancement using low-pass filtering
  and alpha-rooting. Signal Processing  \textbf{170},  107445 (2020).
  \doi{https://doi.org/10.1016/j.sigpro.2019.107445},
  \url{http://www.sciencedirect.com/science/article/pii/S0165168419304967}

\bibitem{fattal_dehazing}
Fattal, R.: Single image dehazing. ACM Trans. Graph.  \textbf{27}(3),
  72:1--72:9 (Aug 2008). \doi{10.1145/1360612.1360671},
  \url{http://doi.acm.org/10.1145/1360612.1360671}

\bibitem{galdran_retinex}
Galdran, A., Bria, A., Alvarez-Gila, A., Vazquez-Corral, J., Bertalmio, M.: On
  the duality between retinex and image dehazing. 2018 IEEE/CVF Conference on
  Computer Vision and Pattern Recognition  (Jun 2018).
  \doi{10.1109/cvpr.2018.00857},
  \url{http://dx.doi.org/10.1109/cvpr.2018.00857}

\bibitem{ietk_github}
Gaudio, A.: Open source code. \url{https://github.com/adgaudio/ietk-ret} (2020)

\bibitem{he_DCP}
{He}, K., {Sun}, J., {Tang}, X.: Single image haze removal using dark channel
  prior. IEEE Transactions on Pattern Analysis and Machine Intelligence
  \textbf{33}(12),  2341--2353 (Dec 2011). \doi{10.1109/TPAMI.2010.168}

\bibitem{he_guided_filter}
{He}, K., {Sun}, J., {Tang}, X.: Guided image filtering. IEEE Transactions on
  Pattern Analysis and Machine Intelligence  \textbf{35}(6),  1397--1409 (June
  2013). \doi{10.1109/TPAMI.2012.213}

\bibitem{dcp_survey}
Lee, S., Yun, S., Nam, J.H., Won, C.S., Jung, S.W.: A review on dark channel
  prior based image dehazing algorithms. EURASIP Journal on Image and Video
  Processing  \textbf{2016}(1), ~4 (Jan 2016). \doi{10.1186/s13640-016-0104-y},
  \url{https://doi.org/10.1186/s13640-016-0104-y}

\bibitem{srinivasa_atmosphere}
Narasimhan, S.G., Nayar, S.K.: Vision and the atmosphere. IJCV  \textbf{48}(3),
   233--254 (January 2002)

\bibitem{book_with_unsharp_mask_algo_p_357}
Petrou, M., Petrou, C.: Image Processing: The Fundamentals, pp. 357--360. John
  Wiley \& Sons, Ltd, Chichester, UK (2011-01-27)

\bibitem{unet_ronneberger}
Ronneberger, O., Fischer, P., Brox, T.: U-net: Convolutional networks for
  biomedical image segmentation. In: Navab, N., Hornegger, J., Wells, W.M.,
  Frangi, A.F. (eds.) Medical Image Computing and Computer-Assisted
  Intervention -- MICCAI 2015. pp. 234--241. Springer International Publishing,
  Cham (2015)

\bibitem{IDRiD}
Sahasrabuddhe, P.P.S.P.R.K.M.K.G.D.V., Meriaudeau, F.: Indian diabetic
  retinopathy image dataset (idrid) (2018). \doi{10.21227/H25W98},
  \url{http://dx.doi.org/10.21227/H25W98}

\bibitem{illumination_dehazing}
{Savelli}, B., {Bria}, A., {Galdran}, A., {Marrocco}, C., {Molinara}, M.,
  {Campilho}, A., {Tortorella}, F.: Illumination correction by dehazing for
  retinal vessel segmentation. In: 2017 IEEE 30th International Symposium on
  Computer-Based Medical Systems (CBMS). pp. 219--224 (June 2017).
  \doi{10.1109/CBMS.2017.28}

\bibitem{smailagic_inverted_dehazing}
Smailagic, A., Sharan, A., Costa, P., Galdran, A., Gaudio, A., Campilho, A.:
  Learned pre-processing for automatic diabetic retinopathy detection on eye
  fundus images. In: Karray, F., Campilho, A., Yu, A. (eds.) Image Analysis and
  Recognition. pp. 362--368. Springer International Publishing, Cham (2019)

\bibitem{tan_visibility_bad_weather}
{Tan}, R.T.: Visibility in bad weather from a single image. In: 2008 IEEE
  Conference on Computer Vision and Pattern Recognition. pp.~1--8 (June 2008).
  \doi{10.1109/CVPR.2008.4587643}

\bibitem{bright_channel_prior}
Wang, Y., Zhuo, S., Tao, D., Bu, J., Li, N.: Automatic local exposure
  correction using bright channel prior for under-exposed images. Signal
  Processing  \textbf{93}(11),  3227 -- 3238 (2013).
  \doi{https://doi.org/10.1016/j.sigpro.2013.04.025},
  \url{http://www.sciencedirect.com/science/article/pii/S0165168413001680}

\end{thebibliography}

\end{document}